\documentclass[conference]{IEEEtran}
\IEEEoverridecommandlockouts
\usepackage{cite}
\usepackage{amsmath,amssymb,amsfonts}
\usepackage{algorithmic}
\usepackage{graphicx}
\usepackage{textcomp}
\usepackage{xcolor}
\usepackage{url}
\usepackage{hyperref}
\usepackage{caption}
\def\BibTeX{{\rm B\kern-.05em{\sc i\kern-.025em b}\kern-.08em
    T\kern-.1667em\lower.7ex\hbox{E}\kern-.125emX}}
\begin{document}

\title{Dimensionality Reduction for Cyberattack Classification: A Comparative Evaluation of PCA and Linear Predictive Coding\\
}

\author{\IEEEauthorblockN{Nelly Elsayed}
\IEEEauthorblockA{\textit{School of Information Technology} \\
\textit{University of Cincinnati}\\
Ohio, United States \\
elsayeny@ucmail.uc.edu}
\and
\IEEEauthorblockN{Zag ElSayed}
\IEEEauthorblockA{\textit{School of Information Technology} \\
\textit{University of Cincinnati}\\
Ohio, United States \\
elsayezs@ucmail.uc.edu}
\and
\IEEEauthorblockN{Navid Asadizanjani}
\IEEEauthorblockA{\textit{Dept. of Electrical \& Computer Engineering} \\
\textit{University of Florida}\\
Florida, United States\\
nasadi@ece.ufl.edu}
}

\maketitle

\begin{abstract}
High-dimensional feature representations are widely used in machine learning-based cyberattack detection systems. However, they increase computational complexity and may hinder deployment in resource-constrained environments. In this paper, we investigate feature compression techniques for cyberattack classification by comparing two dimensionality reduction approaches: Principal Component Analysis (PCA) and Linear Predictive Coding (LPC). Compressed feature representations with varying dimensionalities are generated and evaluated across several classification models. Experimental analysis demonstrates that PCA preserves classification performance even under aggressive compression. On the other hand, LPC provides competitive predictive representations with slightly larger performance degradation. The results show that substantial reductions in feature dimensionality can be achieved with minimal impact on classification accuracy, highlighting the potential of lightweight feature compression for efficient cybersecurity analytics.
\end{abstract}


\section{Introduction}

Machine learning based cyberattack classification models often rely on high-dimensional feature representations to capture complex network behavior~\cite{abdulhammed2019features,hussain2020machine,parizad2022cyber}. These representations can support high predictive accuracy~\cite{zhang2016compact,georgiou2020survey}. However, they increase implementation costs, including memory usage, training costs, and inference latency~\cite{gnana2016literature,ghaddar2018high}. These limitations become more significant in lightweight or resource-constrained environments, including Internet of Things (IoT), wearable devices, Internet of Medical Things (IoMT), edge monitoring systems, and embedded security platforms~\cite{elsayed2025cybersecurity,lone2023comprehensive,elsayed2025cryptodna,uprety2020reinforcement,dzamesi2025review,elsayed2025extreme,al2020survey,aileni2020cybersecurity,elsayed2023iot,ranaweera2021survey}.

Feature compression presents a promising approach to reduce representation size while preserving the discriminative information required for classification~\cite{niu2022exploiting,behiry2024cyberattack}. Principal Component Analysis (PCA) is one of the most widely used dimensionality reduction techniques, which projects data onto a lower-dimensional subspace that preserves the dominant variance structure~\cite{abdi2010principal,karamizadeh2013overview,greenacre2022principal}. Another alternative is Linear Predictive Coding (LPC), which is a signal processing method that represents data using predictive coefficients derived from linear relationships among samples~\cite{spratling2017review}.

This paper presents a comparative evaluation of PCA and LPC for lightweight cyberattack classification. The goal is to determine how strongly feature dimensionality can be reduced while maintaining classification performance across different learning models. 

The contributions of this work are threefold. First, we investigate two feature-compression techniques, Principal Component Analysis (PCA) and Linear Predictive Coding (LPC), for cyberattack classification. Second, we analyze how these methods behave under aggressive dimensionality reduction by evaluating compressed feature representations of four, eight, and 12 dimensions. Third, we conduct a comparative evaluation across multiple classifier families to study the robustness of compressed feature representations and the impact of dimensionality reduction on classification performance.

\section{Background}

\subsection{Principal Component Analysis}
Principal Component Analysis (PCA) is a linear dimensionality reduction technique that transforms correlated variables into a smaller set of uncorrelated variables called principal components~\cite{abdi2010principal}. For a data matrix $X$, PCA computes directions of maximum variance and projects the data onto the leading components:
\begin{equation}
    Z = XV_{k}
\end{equation}
\noindent
where $V_{k}$ contains the top $k$ eigenvectors of the covariance matrix. PCA is widely used in pattern recognition, compression, and anomaly detection because it often preserves useful structure in a compact representation.

\subsection{Linear Predictive Coding}
Linear Predictive Coding (LPC)~\cite{spratling2017review} is a signal processing technique that models a signal sample as a linear combination of previous samples:
\begin{equation}
    x(n) = \sum_{k=1}^{p} a_{k}x(n-k) +e(n)
\end{equation}
\noindent
where $a_{k}$ are the predictive coefficients, $p$ is the LPC order, and e(n) is the prediction error. LPC coefficients are typically estimated using the Levinson-Durbin recursion. LPC has been used extensively in speech coding~\cite{rahmatallah2025pre,madane2009speech}, audio analysis~\cite{grama2017audio,lin1989qrs}, and biomedical signal processing~\cite{carotti2009compression}. In this paper, we use the LPC to derive compressed predictive representations from cybersecurity feature vectors.

\section{Methodology}

\begin{figure}[htbp]
\centerline{\includegraphics[width=1.0\columnwidth]{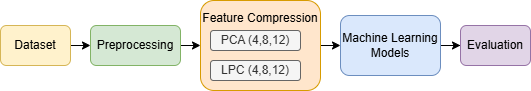}}
\caption{Experimental pipeline for feature compression, classification, and evaluation.}
\label{fig:pipeline}
\end{figure}

\subsection{Experimental Pipeline}

The overall experimental workflow is shown in Fig.~\ref{fig:pipeline}. The process begins with data preprocessing, where multiple CSV files are merged into a unified dataset and cleaned to remove missing or invalid values. The resulting feature vectors are then used to generate compressed representations using two dimensionality reduction techniques: Principal Component Analysis (PCA) and Linear Predictive Coding (LPC).

These compressed representations are subsequently used as inputs to multiple machine learning classifiers. All models are trained and evaluated using the same training and testing partitions to ensure a consistent comparison between feature representations. Performance is measured using accuracy, weighted recall, and weighted F1-score.

\subsection{Feature Compression}

Two dimensionality reduction techniques are investigated in this work: Principal Component Analysis (PCA) and Linear Predictive Coding (LPC). PCA reduces the dimensionality of the original feature space by projecting the data onto a set of orthogonal components that capture the dataset's largest variance. By retaining only the leading components, PCA produces a compact representation that preserves the data's dominant structure.

LPC provides an alternative compression mechanism based on predictive modeling. Unlike PCA, which preserves directions of maximum variance, LPC represents a feature vector using a compact set of predictive coefficients. In this work, LPC is applied directly to the ordered cybersecurity feature vector associated with each network flow. Let $\mathbf{x}=[x_1,x_2,\ldots,x_d]$ denote the original feature representation. LPC estimates a set of coefficients that model each element as a linear combination of preceding elements in the ordered vector, producing a compressed representation of dimensionality $p$.

It is important to note that LPC is not used here to model packet-level temporal behavior. Rather, it is investigated as a predictive coding technique for compact representation learning in tabular cybersecurity data. The original feature ordering provided by the dataset is retained during LPC extraction, allowing the predictive coefficients to summarize dependencies among neighboring feature dimensions. The objective is not to impose a temporal interpretation on the data, but instead to evaluate whether predictive coefficient representations can preserve discriminative information under aggressive dimensionality reduction.

To analyze the effect of aggressive dimensionality reduction, compressed feature representations of four, eight, and 12 dimensions are generated using both methods.

\begin{table}[t]
\caption{Feature dimensionality reduction achieved by compressed representations}
\begin{center}
\begin{tabular}{|c|c|c|}
\hline
\textbf{Feature Representation} & \textbf{Number of Features} & \textbf{Reduction (\%)} \\
\hline
Raw & 78 & 0 \\
PCA-12 & 12 & 84.6 \\
PCA-8 & 8 & 89.7 \\
PCA-4 & 4 & 94.9 \\
LPC-12 & 12 & 84.6 \\
LPC-8 & 8 & 89.7 \\
LPC-4 & 4 & 94.9 \\
\hline
\end{tabular}
\label{tab:reduction}
\end{center}
\end{table}

Table~\ref{tab:reduction} summarizes the dimensionality reduction achieved by the compressed feature representations. Both PCA and LPC significantly reduce the size of the input feature space. In particular, the PCA-4 and LPC-4 representations reduce the dimensionality by approximately 94.9\% compared to the original feature space.

\subsection{Classification Models}

To evaluate the robustness of compressed feature representations, we employed multiple machine learning classifiers. Logistic Regression and Linear Support Vector Machines are included as baseline linear models. Random Forest and Gradient Boosting are used as ensemble tree-based methods capable of capturing nonlinear feature interactions. In addition, a Multi-Layer Perceptron (MLP) neural network is used to represent a deep learning-based classifier.

Evaluating multiple model families enables us to examine how different learning paradigms respond to aggressive feature compression and to identify which models remain most robust under reduced feature dimensionality.

\subsection{Dataset}

Experiments are conducted using the CICIDS2017 (Canadian Institute for Cybersecurity Intrusion Detection System 2017) dataset~\cite{sharafaldin2018toward,unb_cicids2017}, a widely used benchmark for evaluating intrusion detection systems. The dataset contains labeled network traffic flows representing both benign activity and multiple cyberattack categories.

After merging the dataset files and removing records containing missing or invalid values, the resulting dataset contains 2,574,264 samples described by 78 numerical features. The traffic flows belong to 15 classes, including benign traffic and several attack types such as denial-of-service, distributed denial-of-service, port scanning, brute-force attacks, and infiltration activities. The dataset is split into 80\% training and 20\% testing subsets using stratified sampling to preserve the class distribution during model evaluation.

\subsection{Model Implementation}

All experiments are implemented in Python using commonly used machine learning libraries. Data preprocessing and feature manipulation are performed using Pandas and NumPy. PCA-based dimensionality reduction and the classical machine learning classifiers are implemented using Scikit-learn.

The neural network classifier is implemented in PyTorch using a Multi-Layer Perceptron architecture composed of fully connected layers with nonlinear activation functions. Training is performed using the Adam optimizer with mini-batch gradient descent. Early stopping based on validation performance is applied to mitigate overfitting.

Model performance is evaluated using accuracy, weighted recall, weighted F1-score, and Macro-F1. Weighted F1-score is used for comparing feature representations, whereas Macro-F1 is reported to provide additional insight into class-wise performance.

\begin{figure}[t]
\centerline{\includegraphics[width=7.5cm, height = 4 cm]{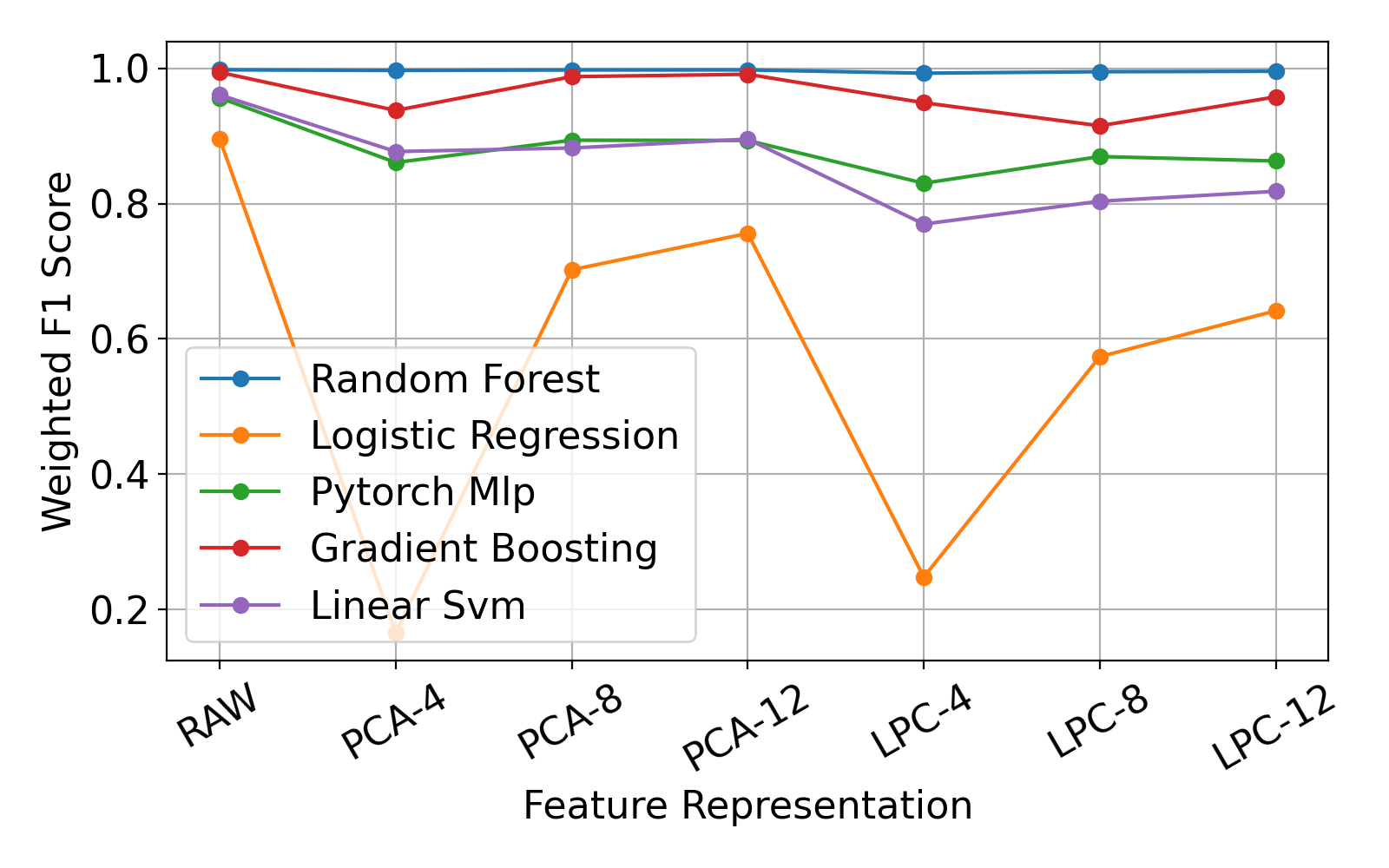}}
\caption{Classifier performance across feature representations.}
\label{fig:models_compare}
\end{figure}

\begin{table}[t]
\caption{Best performance for each feature representation}
\begin{center}
\begin{tabular}{|p{3cm}|p{1.2cm}|p{2.5cm}|}
\hline
\textbf{Feature Representation} & \textbf{Dimension}& \textbf{Weighted F1-score} \\
\hline
Raw	&78	&0.9981\\
\hline
PCA-4&	4&	0.9972\\
\hline
PCA-8	&8&	0.9979\\
\hline
PCA-12&	12&	0.9980\\
\hline
LPC-4	&4	&	0.9931\\
\hline
LPC-8&	8	&0.9954\\
\hline
LPC-12&	12	&0.9960\\
\hline
\end{tabular}
\label{tab1}
\end{center}
\end{table}

\begin{figure}[t]
\centerline{\includegraphics[width=7cm, height = 4 cm]{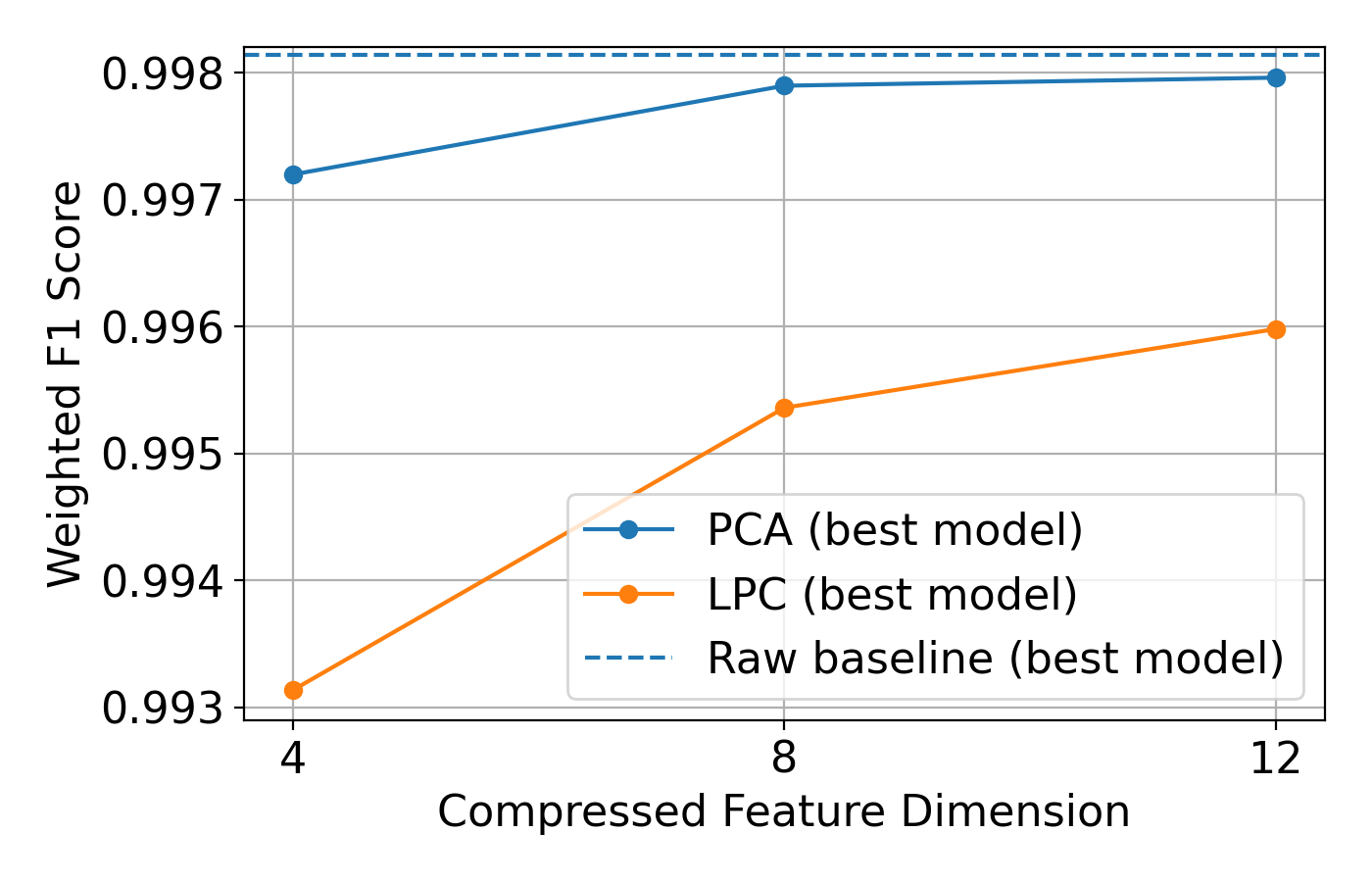}}
\caption{Compression–performance relationship for PCA and LPC.}
\label{fig:Compare_diagram}
\end{figure}

\begin{figure}[t]\vspace{-10pt}
\centerline{\includegraphics[width=7cm, height = 4 cm]{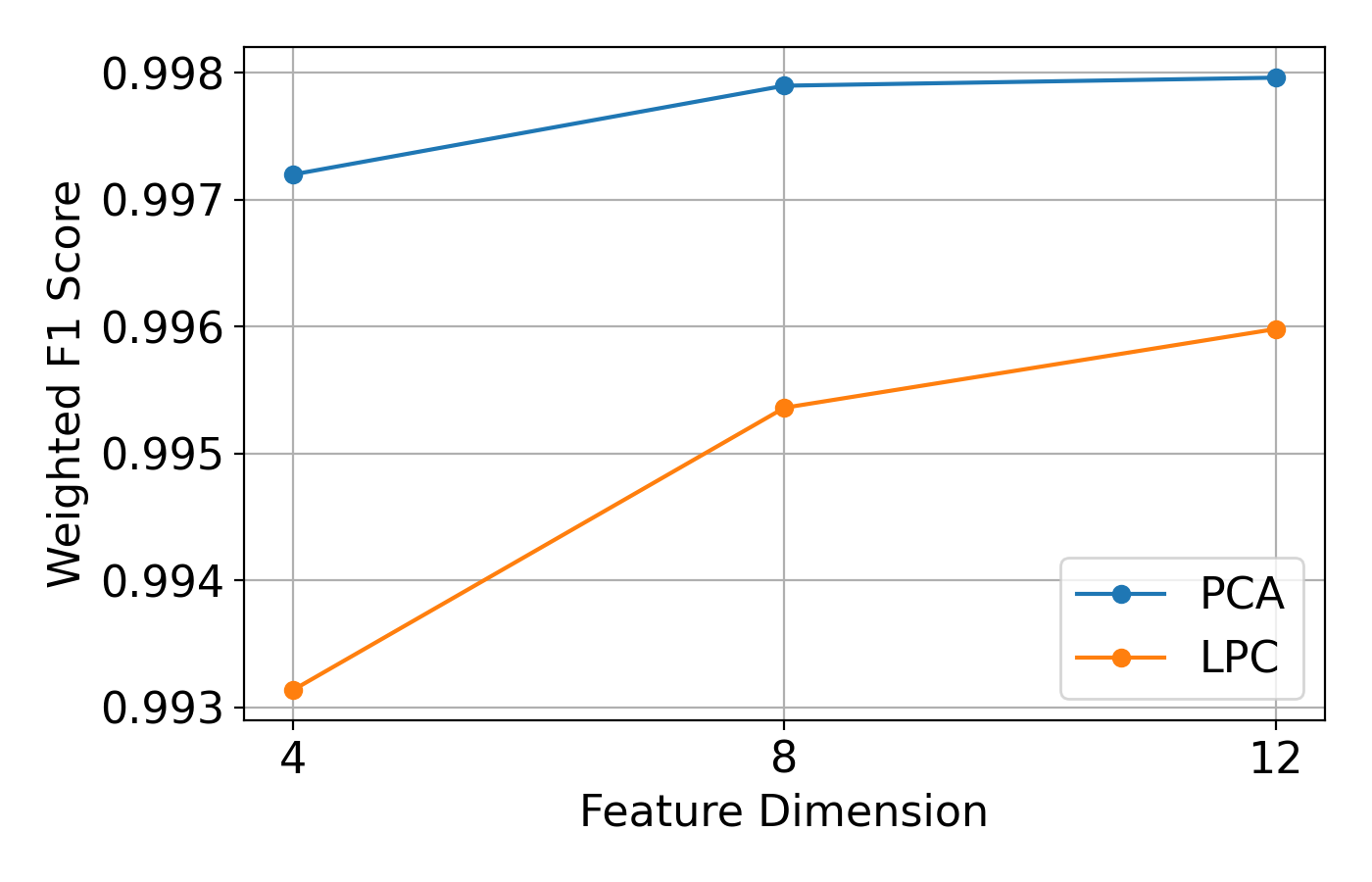}}
\caption{Comparison of PCA and LPC feature representations at different compression levels.}
\label{fig:pca_vs_lpc}
\end{figure}

\section{Results and Discussion}

\subsection{Classifier Comparison}

The performance of all evaluated classifiers across the different feature representations is shown in Fig.~\ref{fig:models_compare}. The figure compares the weighted F1-scores obtained by Logistic Regression, Linear Support Vector Machines, Random Forest, Gradient Boosting, and a Multi-Layer Perceptron (MLP).

As observed in Fig.~\ref{fig:models_compare}, Random Forest consistently achieves the highest performance across all feature representations. Gradient Boosting also performs strongly but remains slightly below Random Forest in most cases. In contrast, linear models such as Logistic Regression and Linear SVM exhibit larger performance degradation when aggressive dimensionality reduction is applied. The MLP model achieves competitive results but does not outperform ensemble tree-based methods.

These results indicate that ensemble tree-based classifiers are more robust to feature compression. Since Random Forest consistently performs best, it is used as the primary reference model for analyzing the impact of dimensionality reduction in the following subsection.

\subsection{Feature Compression Analysis}

Table~\ref{tab1} summarizes the best classification performance obtained for each feature representation, while Table~\ref{tab:macrof1} reports Macro-F1 scores for the strongest Random Forest configurations to provide additional insight into class-wise performance. Even under aggressive dimensionality reduction, the compressed feature sets maintain performance close to the original representation. Notably, PCA-4 reduces the feature dimensionality by approximately 94.9\% while maintaining nearly identical classification performance compared to the full feature set.

The results demonstrate that substantial dimensionality reduction can be achieved with only a minor loss in classification performance, particularly when PCA is used.

\begin{table}[t]
\centering
\caption{Comparison of Random Forest performance using Accuracy and Macro-F1.}
\label{tab:macrof1}
\begin{tabular}{lcc}
\hline
\textbf{Method} & \textbf{Accuracy} & \textbf{Macro-F1} \\
\hline
Raw Features      & \textbf{0.9982} & \textbf{0.9055} \\
PCA-8             & 0.9979 & 0.7856 \\
PCA-4             & 0.9973 & 0.7852 \\
LPC-8             & 0.9954 & 0.7739 \\
LPC-4             & 0.9932 & 0.7651 \\
LPC-12            & 0.9961 & 0.7575 \\
\hline
\end{tabular}
\end{table}

\begin{table}[t]
\centering
\caption{Computational cost of Random Forest using selected feature representations.}
\label{tab:runtime}
\begin{tabular}{lccc}
\hline
\textbf{Representation} & \textbf{Features} & \textbf{Train Time (s)} & \textbf{Test Time (s)}\\
\hline
Raw & 78 & 290.12 & 4.14\\
PCA-8 & 8 & 221.84 & 3.66\\
LPC-8 & 8 & 300.53 & 3.85\\
\hline
\end{tabular}
\end{table}

To provide a more balanced evaluation under the severe class imbalance present in CICIDS2017, Macro-F1 scores were additionally analyzed for the best-performing Random Forest classifier. While Random Forest maintained classification accuracies above 99\% across all compressed feature representations, Macro-F1 revealed larger differences in class-wise detection performance. The Random Forest classifier operating on the original feature space achieved the highest Macro-F1 score of 0.9055. Among the compressed representations, PCA-based feature reduction preserved minority-class information more effectively than LPC. The best PCA configuration (PCA-8) achieved a Macro-F1 score of 0.7856, whereas the best LPC configuration (LPC-8) achieved 0.7739.

Table~\ref{tab:runtime} reports the computational cost of the Random Forest classifier using the original and compressed feature representations. PCA-8 reduces the feature dimensionality from 78 to 8 features and decreases both training and testing time while maintaining high classification performance. LPC-8 also reduces the input dimensionality to 8 features and achieves lower testing time than the raw feature representation, although its coefficient-based representation results in slightly higher training time. These results support the use of compact feature representations for lightweight cybersecurity analytics.

The relationship between feature dimensionality and classification performance is shown in Fig.~\ref{fig:Compare_diagram}. PCA preserves performance even under strong dimensionality reduction. Reducing the feature space from 78 features to only four principal components results in only a minor decrease in weighted F1-score. 

LPC also provides effective dimensionality reduction, although the performance degradation is slightly larger at lower dimensions. Increasing the LPC order improves classification performance, indicating that higher-order predictive representations retain more information from the original feature vectors.

A direct comparison between PCA and LPC is shown in Fig.~\ref{fig:pca_vs_lpc}. At equivalent dimensionalities, PCA consistently produces slightly higher classification performance. Nevertheless, LPC maintains strong results and demonstrates that predictive compression techniques can provide compact feature representations while preserving most of the discriminative information required for classification. Overall, the results show that substantial feature compression can be achieved with minimal loss in classification accuracy. These findings highlight the potential of lightweight feature representations for efficient machine learning-based cybersecurity analytics.

\begin{table}[t]
\centering
\caption{Class-wise F1-scores for selected attack categories using Random Forest with raw features.}
\label{tab:lowest}
\begin{tabular}{lc}
\hline
\textbf{Attack Class} & \textbf{F1-score }\\
\hline
Bot & 0.8163 \\
Infiltration & 0.8333 \\
Heartbleed & 1.0000 \\
Web Attack--Brute Force & 0.7752 \\
Web Attack--SQL Injection & 0.8571 \\
Web Attack--XSS & 0.3364 \\
\hline
\end{tabular}
\end{table}

Table~\ref{tab:lowest} reports class-wise F1-scores for selected attack categories. Most attack types are detected with high reliability, including Heartbleed, Infiltration, Bot, and Web Attack-SQL Injection. Although performance varies across attack categories, the results indicate that the Random Forest classifier maintains strong detection capability across a diverse set of cyberattack types.

The observed performance differences between PCA and LPC can be attributed to the underlying mechanisms of the two compression techniques. PCA preserves the directions of maximum variance in the feature space, which often correspond to the most informative components for classification tasks. As a result, even highly compressed PCA representations retain much of the discriminative structure of the original data. In contrast, LPC derives predictive coefficients based on linear relationships among feature elements. While this representation effectively captures structural dependencies within the feature vector, it does not explicitly preserve variance across the entire feature space, which may lead to slightly larger performance degradation under aggressive compression.

From a practical perspective, reducing the dimensionality of cybersecurity feature representations also improves computational efficiency during both training and inference. Lower-dimensional inputs require fewer arithmetic operations and reduce memory usage, which can be particularly beneficial for deployment in resource-constrained environments such as IoT devices, edge monitoring systems, and embedded security platforms. Therefore, lightweight feature compression techniques such as PCA and LPC can play an important role in enabling scalable and efficient machine learning-based cybersecurity analytics in real-world systems.

\section{Conclusion}
This paper presented a comparative study of two feature compression techniques, Principal Component Analysis (PCA) and Linear Predictive Coding (LPC), for lightweight cyberattack classification. The goal was to analyze how aggressively feature dimensionality can be reduced while maintaining high classification performance across different machine learning models. 

Our experimental results showed that ensemble tree-based classifiers, particularly Random Forest, consistently achieved the strongest performance across all feature representations. PCA demonstrated the best trade-off between compression and performance, preserving near-baseline classification accuracy even under substantial dimensionality reduction. LPC also produced competitive compressed representations, although with slightly larger performance degradation at lower dimensions. 

These findings show that predictive signal-processing techniques, such as LPC, can provide compact yet informative representations of cybersecurity data, suggesting that signal modeling approaches offer a promising direction for lightweight, efficient cyberattack detection.

\bibliographystyle{ieeetr}
\bibliography{references}

\end{document}